\documentclass[12pt,a4paper]{article}
\usepackage{amsmath}
\usepackage{amsfonts}
\usepackage{amssymb}
\usepackage{textcomp}
\usepackage[dvips]{graphicx}
\usepackage{epsfig}
\usepackage{bm}
\usepackage{dcolumn}
\usepackage{amsfonts}
\usepackage{color}
\usepackage[left=2cm,right=2cm,top=2cm,bottom=2cm]{geometry}
\usepackage[left=2cm,right=2cm,top=2cm,bottom=2cm]{geometry}
\usepackage[tableposition=top]{caption}
\usepackage{subcaption}
\DeclareCaptionLabelFormat{gostfigure}{Fig. #2}
\DeclareCaptionLabelSeparator{gost}{.~~}
\begin{document}
\title{\textbf{Bouncing solutions in $f(T)$ gravity}}
\author{Maria A. Skugoreva$^{1}$\footnote{masha-sk@mail.ru}, Alexey V. Toporensky$^{1,2}$\footnote{atopor@rambler.ru}\vspace*{3mm} \\
\small $^{1}$Kazan Federal University, Kremlevskaya 18, Kazan 420008, Russia\\
\small $^{2}$Sternberg Astronomical Institute, Lomonosov Moscow State University,\\
\small Moscow 119991, Russia}

\date{ \ }

\maketitle

\begin{abstract}
    We consider certain aspects of cosmological dynamics of a spatially curved Universe in $f(T)$ gravity. Local analysis allows us to find conditions for bounces and for static solutions; these conditions appear to be in general less restrictive than in general relativity. We also provide a global analysis of the corresponding cosmological dynamics in the cases when bounces and static configurations exist, by constructing phase diagrams. These diagrams indicate that the fate of a big contracting Universe is not altered significantly when bounces become possible, since they appear to be inaccessible by a sufficiently big Universe.
\end{abstract}

\section{Introduction}
~~~~Modified gravity can lead to some kinds of cosmological dynamics which are impossible in general relativity (GR), at least for usual matter content of the Universe, like a perfect fluid
with positive energy density. One of well-known examples is the so-called non-standard singularity
where the scale factor $a$, Hubble parameter $H$ and the matter energy density $\rho$ remain constant, while $\dot H$ diverges \cite{Barrow}. Evolution of the Universe cannot be prolonged through this point.

    A non-zero spatial curvature gives even more diversity in possible dynamical regimes. We should remind the reader that in GR the influence of the spatial curvature upon the cosmological dynamics of an isotopic Universe filled with a perfect fluid is quite easy to explain. The Friedmann equation contains only three terms:
$$
\frac{3}{8\pi}{M_{Pl}}^2\left( H^2 + \frac{k}{a^2}\right) = \rho,
$$
where $k=0, 1, -1$ for zero, positive and negative spatial curvature, respectively. Qualitative features of the dynamics are determined completely if we know which term (with the curvature or the matter energy density) dominates at the particular epoch. Since the energy density of a perfect fluid with the equation of state parameter $w$ falls as $a^{2/[3(w+1)]}$, the matter energy density dominates at small $a$ for $w>-1/3$ and for large $a$ in the opposite case of $w<-1/3$. For positive spatial curvature this leads to an ultimate bounce in the case of $w<-1/3$ and an ultimate recollapse for $w>-1/3$. A particular value of the energy density of matter makes possible a static solution in the $w<-1/3$ case which is known to be unstable. On the contrary, negative spatial curvature does not change the general evolution of the Universe from Big Bang to an eternal expansion, and the only difference between two cases, $w>-1/3$ and $w<-1/3$ is the time location of curvature dominated Milne asymptotic $a\propto t$, which occurs near Big Bang for $w<-1/3$ and at late times for $w>-1/3$.

    In modified gravity the equations of motion contain more terms, and the effects of curvature cannot be described so easily. Moreover, unlike GR where the sign of the curvature term in the equation of motion is fixed by the sign of the curvature parameter $k$, the signs of the curvature containing terms in the equations of motion of modified gravity can be different, resulting in possible complicated dynamics for a negative spatial curvature as well. 

    An interesting example of a cosmological solution in modified gravity for a curved Universe impossible in GR is a stable static Universe in $f(R)$ gravity \cite{Dunsby} and in Gauss-Bonnet gravity \cite{Lobo} and massive gravity \cite{Parisi}. Later, similar solutions have been found in $f(T)$ gravity for quadratic \cite{Wu} and exponential functions $f(T)$ \cite{Li}.
    
     The $f(T)$ theory became an area of intense investigation less than a decade ago. Instead of the metric the fundamental objects of this theory are tetrads, and instead of a torsion free Levi-Civita connection it uses curvature free connections (for example, Weitzenb\"{o}ck connection~\cite{Weitzenbock}). The equations of motion in GR can be obtained in torsion language resulting in so-called teleparallel equivalent of general relativity (TEGR) (see the book of Ref.~\cite{Pereira}). Then $f(T)$ gravity \cite{Cai} can be obtained from TEGR in the same way as $f(R)$ theory generalizes GR. However, in contrast to the relation between TEGR and GR the dynamical equations in $f(T)$ \cite{Cai, FF, Ferraro1, Linder} and $f(R)$ \cite{Sotiriou} are different. A nice feature of the equations of motion in $f(T)$ gravity is that they are second-order differential equations in contrast to forth-order equations in $f(R)$ gravity. However, the number of degrees of freedom in $f(T)$ gravity is still unknown \cite{Guzman} and is subject to ongoing debates. The reason is the lack of local Lorentz invariance \cite{Lorentz1, Lorentz2} and in general it forces one to use a particular tetrad in order to get the correct equations of motion. There are several methods of finding this so-called ``proper'' tetrad \cite{Obukhov, Ferraro2, Tamanini, Ferraro3, Martin, Jarv}; however, no one seems to be universal. However, treatment of a system with higher symmetries is easier, and the Friedmann-Lema\^{i}tre-Robertson-Walker (FLRW) Universe belongs to this class. It is known that Cartesian tetrad is the proper one for a flat FLRW Universe, and corresponding cosmological dynamics has been studied in detail, including a comparison with observations \cite{Sar, we, we1, Masha, DA1, DA2, Na, Ubossym}. As for a spatially curved Universe, the correct proper tetrad has been found in \cite{Ferraro2}. Using this tetrad it is possible to write down the corresponding equations of motion. For a detailed analysis of these equations, see, for example, \cite{Ferraro4, Capozziello}. 
    
    In the present paper we extend the analysis of the static Universe provided in \cite{Wu} to a general power-law $f(T)$. Another pecularity of modified gravity is the possibility of bounce for a larger range of the equation of state parameter $w$ than in GR for the positive spatial curvature, and bounces in the case of the negative spatial curvature. The existence of stable static solutions implies bounces (see below); however, the conditions for the bounce to exist are in general softer. Particular forms of bouncing solutions have been considered earlier for the case of a flat metric. However, a bounce in a flat Universe requires functions $f(T)$ either with $f(0) \ne 0$ \cite{Cai1}| --- or with $\frac{df}{dT}\big{\vert}_{T=0}$ diverging \cite{Bamba}. None of these conditions can be realized for a power-law $f(T)$ with integer index. Other possibilities require a theory beyond $f(T)$ \cite{Jackson1, Jackson2}. In the present paper we write down general  conditions for a bouncing spatially curved Universe in power-law $f(T)$ gravity and construct the corresponding phase diagrams.
    
    The structure of the paper is as follows: in Sect.~\ref{sec2} we remind the reader proper tetrads and the equations of motion for a spatially curved isotropic Universe. In Sect.~\ref{sec3} using these equations we consider static solutions and their stability, as well as local conditions for a bounce. In Sect.~\ref{sec4} we construct the phase diagrams and discuss some features of global dynamics which can be missed in local analysis. Section~\ref{sec5} provides a summary of the results obtained.

\section{The equations of motion}
\label{sec2}
~~~~In the present paper we consider the cosmological models in $f(T)$ gravity with matter. The action of this theory is
\begin{equation}
\label{action}
S=\frac{1}{2K}\int e~ f(T)d^4x +S_m,
\end{equation}
where $e=\text{det}(e^A_{\mu})=\sqrt{-g}$ is the determinant, which consists of the tetrad components $e^A_{\mu}$, ~~$f(T)$ is a general differentiable function of the torsion scalar $T$, ~~$S_m$ is the matter action and $K=8\pi G$. Here the units $\hbar=c=1$ are used.  

    The line element of a non-flat homogeneous and isotropic Friedmann-Lema\^{i}tre-Robertson-Walker Universe is
\begin{equation}
\label{metric}
ds^2=dt^2-{\varkappa}^2a^2(t)\Big{[}d{(\varkappa\psi)}^2+{\sin}^2(\varkappa\psi)(d{\theta}^2+{\sin}^2\theta d{\varphi}^2)\Big{]},
\end{equation}
where ~~$a(t)$~~ is the scale factor, the parameter ~~$\varkappa=1$~~ for ta closed Universe and ~~$\varkappa=i$~~ for an open Universe.

    We write down the following diagonal tetrad (see \cite{Ferraro2}, \cite{Ferraro4}), which relates to the metric (\ref{metric}): 
\begin{equation}
\label{tetrad}
e^A_\mu =(1, a(t)E^1, a(t)E^2, a(t)E^3),
\end{equation}    
where for $\varkappa=1$
\begin{equation}
\begin{array}{l}
\label{Ek1}
E^1=-\cos\theta d\psi+\sin\psi\sin\theta(\cos\psi d\theta-\sin\psi\sin\theta d\varphi),\\
\\E^2=\sin\theta\cos\varphi d\psi-\\
-\sin\psi\Big{[}(\sin\psi\sin\varphi-\cos\psi\cos\theta\cos\varphi)d\theta+(\cos\psi\sin\varphi+\sin\psi\cos\theta\cos\varphi)\sin\theta d\varphi\Big{]},\\
\\E^3=-\sin\theta\sin\varphi d\psi-\\
-\sin\psi\Big{[}(\sin\psi\cos\varphi+\cos\psi\cos\theta\sin\varphi)d\theta+(\cos\psi\cos\varphi-\sin\psi\cos\theta\sin\varphi)\sin\theta d\varphi\Big{]}
\end{array}
\end{equation}
and for $\varkappa=i$
\begin{equation}
\begin{array}{l}
\label{Ek2}
E^1=\cos\theta d\psi+\sinh\psi\sin\theta(-\cosh\psi d\theta+i\sinh\psi\sin\theta d\varphi),\\
\\E^2=-\sin\theta\cos\varphi d\psi+\\
+\sinh\psi\Big{[}(i\sinh\psi\sin\varphi-\cosh\psi\cos\theta\cos\varphi)d\theta+(\cosh\psi\sin\varphi+i\sinh\psi\cos\theta\cos\varphi)\sin\theta d\varphi\Big{]},\\
\\E^3=\sin\theta\sin\varphi d\psi+\\
+\sinh\psi\Big{[}(i\sinh\psi\cos\varphi+\cosh\psi\cos\theta\sin\varphi)d\theta+(\cosh\psi\cos\varphi-i\sinh\psi\cos\theta\sin\varphi)\sin\theta d\varphi\Big{]}.
\end{array}
\end{equation}

    In what follows we will use the more convenient curvature parameter $k$, such that $k=1$ for a closed Universe and $k=-1$ for an open Universe. The torsion scalar for the chosen tetrad (\ref{tetrad}) is
\begin{equation}
\label{THak}
T=-6H^2+\frac{6k}{a^2},
\end{equation}
where $H_a\equiv\frac{\dot a}{a}$ is the Hubble parameter, a dot denotes the derivative with respect to time. Then the time derivative of the torsion scalar has the form
\begin{equation}
\label{TtHakt}
\dot T=-12\dot H H-\frac{12kH}{a^2}.
\end{equation}

     The equations of motion can be found by varying the action (\ref{action}) with respect to the chosen tetrad (\ref{tetrad}) (see, \cite{Ferraro4}, \cite{Capozziello})
\begin{equation}
\label{constraint}
12H^2f_T+f(T)=2 K\rho,
\end{equation}
\begin{equation}
\label{Ht}
4\left( \dot H+\frac{k}{a^2}\right) (12H^2f_{TT}+f_T)-f(T)-4f_T(2\dot H+3H^2)=2Kw\rho,
\end{equation}
where $\rho$ is the matter energy density, $p$ is its pressure, the equation of state is $p=w\rho$ where $w$ is a constant. In the present paper we consider only usual matter, so the equation of state parameter is bounded within the region of thermodynamical stability of the matter $w\in[-1; 1]$. Here we denote $f_T=\frac{d f(T)}{d T}$, ~~ $f_{TT}=\frac{d^2 f(T)}{dT^2}$. 

    The continuity equation for matter is
\begin{equation}
\label{continuity}
\dot\rho+3H(1+w)\rho=0.
\end{equation}

    For the case of $f(T)=T+f_0T^N$ the equations of motion (\ref{constraint}), (\ref{Ht}) have the following form:
\begin{equation}
\label{constraint1}
12H^2(1+Nf_0T^{N-1})+T+f_0T^N=2 K\rho,
\end{equation}
\begin{equation}
\begin{array}{l}
\label{Ht1}
4\left( \dot H+\frac{k}{a^2}\right) (12H^2N(N-1)f_0T^{N-2}+1+Nf_0T^{N-1})-\\
-T-f_0T^N-4(1+Nf_0T^{N-1})(2\dot H+3H^2)=2Kw\rho.
\end{array}
\end{equation}  
In the present paper we consider only integer $N>1$. Equation (\ref{constraint1}) tells us now that a bounce cannot occur in a flat Universe with ordinary matter, since for the flat metric $T=0$ at the point of bounce, and regularity of $f_T(0)$ indicates that left-hand side of Eq.~(\ref{constraint1}) vanishes at the bounce, which is incompatible with positivity of $\rho$.
 
\section{Bouncing solutions and a static Universe}
\label{sec3}
~~~~Looking at Eq.~(\ref{constraint1}) we can see two important points. First, there are curvature-containing terms, originating from the last term in the right-hand side of (\ref{constraint1}) which grow faster than $1/a^2$ near the singularity $a=0$. The fastest term is equal to $(6 k)^N/a^{2 N}$, and this is the only correction term which does not vanish at a bounce. Thus, this term drives the bounce which we study in the present paper. Note that we need both $f(T)$ corrections and a spatial curvature $k \neq 0$ for this type of bounce. The change of power index from $2$ for GR curvature term to $2N$ for the curvature $f(T)$-corrections term leads to the following: already in quadratic gravity the influence of the corrections of the curvature term may overcome the influence of matter near a singularity for $w<1/3$ instead of $w<-1/3$ in GR, and starting from $N=3$ this property covers the whole allowed range $w \in [-1; 1]$. This can induce bounces for a wider range of $w$ than in GR; however, since there are also other terms in (\ref{constraint1}), this possibility is not satisfied automatically and needs further analysis. Second, the sign of the fastest growing curvature term is equal to $k^N$, so it is the same for positive and negative curvature if $N$ is even. This means that we can expect bouncing solutions in quadratic gravity even for $k=-1$. Motivated by this qualitative considerations, we start our analysis by analytical methods. In the next section we present results of numerical investigations.

    We can express $\dot H$ from the equation of motion (\ref{Ht1}) for the model $f(T)=T+f_0T^N$:
\begin{equation}
\begin{array}{l}
\label{Ht2}
4\dot H\Big{(}12H^2N(N-1)f_0T^{N-2}-Nf_0T^{N-1}-1\Big{)}=T+f_0T^N+12H^2(1+Nf_0T^{N-1})-\\
-4\frac{k}{a^2}\Big{(}12H^2N(N-1)f_0T^{N-2}+Nf_0T^{N-1}+1\Big{)}+2Kw\rho.
\end{array}
\end{equation}       
Taking into account (\ref{THak}) and (\ref{constraint1}) we find
\begin{equation}
\label{Ht3}
\dot H=\frac{\frac{K}{2}(w+1)\rho-\frac{k}{a^2}\left[ 12H^2N(N-1)f_0{\left(-6H^2+\frac{6k}{a^2}\right)}^{N-2}+Nf_0{\left(-6H^2+\frac{6k}{a^2}\right)}^{N-1}+1\right] }{12H^2N(N-1)f_0{\left(-6H^2+\frac{6k}{a^2}\right)}^{N-2}-Nf_0{\left(-6H^2+\frac{6k}{a^2}\right)}^{N-1}-1}.
\end{equation}    
From this equation we can finally exclude $\rho$ using (\ref{constraint1}) 
\begin{equation}
\begin{array}{l}
\label{Ht4}
\dot H=\frac{\frac{1}{4}(w+1)\left[ 6H^2+12H^2Nf_0{\left(-6H^2+\frac{6k}{a^2} \right) }^{N-1}+\frac{6k}{a^2}+f_0{\left(-6H^2+\frac{6k}{a^2} \right) }^N\right]}{12H^2N(N-1)f_0{\left(-6H^2+\frac{6k}{a^2}\right)}^{N-2}-Nf_0{\left(-6H^2+\frac{6k}{a^2}\right)}^{N-1}-1}-\\
-\frac{k}{a^2}\frac{ 12H^2N(N-1)f_0{\left(-6H^2+\frac{6k}{a^2}\right)}^{N-2}+Nf_0{\left(-6H^2+\frac{6k}{a^2}\right)}^{N-1}+1 }{12H^2N(N-1)f_0{\left(-6H^2+\frac{6k}{a^2}\right)}^{N-2}-Nf_0{\left(-6H^2+\frac{6k}{a^2}\right)}^{N-1}-1}.
\end{array}
\end{equation}
The equation for $\dot a$ is obviously
\begin{equation}
\label{at}
\dot a=aH.
\end{equation}
The dynamical system (\ref{Ht4}), (\ref{at}) has one stationary point with the following coordinates: 
\\
\\$a=a_0=\sqrt{6k{\left( \frac{f_0(2N-3(w+1))}{3w+1}\right) }^{\frac{1}{N-1}}}$,~~ $H=0$.
\\
It follows from the expression for the coordinate $a_0$ that this point exists for 
\begin{equation}
\label{a0exist1}
6k{\left( \frac{f_0(2N-3(w+1))}{3w+1}\right) }^{\frac{1}{N-1}}\geqslant 0~~\Rightarrow~~
\left[
\begin{array}{l}
\frac{f_0(2N-3(w+1))}{3w+1}\geqslant 0 \text{~~~~for ~~$k=1$},\\
\frac{f_0(2N-3(w+1))}{3w+1}\leqslant 0 \text{~~~~for ~~$k=-1$, $N$ is even.}
\end{array}
\right.
\end{equation}
We substitute the coordinates of the fixed point $H=0$, $a=a_0$ to the constraint equation (\ref{constraint1}) and taking into account positivity of the matter energy density $\rho\geqslant 0$ it is easy to get
\begin{equation}
\begin{array}{l}
\label{a0exist2}
\frac{6k}{{a_0}^2}\left[1+f_0{\left( \frac{6k}{{a_0}^2}\right)}^{N-1}\right]=2K\rho\geqslant0~~\Rightarrow~~ 
\left[
\begin{array}{l}
f_0{\left( \frac{6k}{{a_0}^2}\right) }^{N-1}\geqslant -1 \text{~~~~for ~~$k=1$},\\
f_0{\left( \frac{6k}{{a_0}^2}\right) }^{N-1}\leqslant -1 \text{~~~~for ~~$k=-1$}
\end{array}
\right.
~~\Rightarrow\\
\\\Rightarrow~~
\left[
\begin{array}{l}
\frac{3w+1}{2N-3(w+1)}\geqslant -1 \text{~~~~for ~~$k=1$},\\
\frac{3w+1}{2N-3(w+1)}\leqslant -1 \text{~~~~for ~~$k=-1$}
\end{array}
\right.
~~\Rightarrow~~ 
\left[
\begin{array}{l}
w<\frac{2N-3}{3}, \text{~~~~for ~~$k=1$},\\
w>\frac{2N-3}{3}, \text{~~~~for ~~$k=-1$}.
\end{array}
\right.
\end{array}
\end{equation}
Uniting conditions (\ref{a0exist1}), (\ref{a0exist2}) we find finally that there are three different cases for which the static cosmological solution exists:
\\\textbf{1.} $k=1$,~~ $f_0>0$,~~ $w\in\left(-\frac{1}{3}; \frac{2N-3}{3}\right)$,
\\\textbf{2.} $k=1$,~~ $f_0<0$,~~ $w\in\left[-1; -\frac{1}{3}\right)$,
\\\textbf{3.} $k=-1$,~~ $N$ is even,~~ $f_0>0$,~~ $w\in\left(\frac{2N-3}{3}; 1 \right]$.

    Note that the negative curvature case is realized only for $N=2$ with $w \in (1/3; 1]$. Formally, similar solutions can exist for other even power indices, however, this requires an exotic matter with $w>1$.

    The eigenvalues for the Jacobian matrix associated with the system (\ref{Ht4}), (\ref{at}) in this critical point are
\begin{equation}
\label{lambda}
\lambda_{1,2}=\pm \frac{\sqrt{2}}{6}\sqrt{-\frac{\Big{(}2N-3(w+1)\Big{)}(3w+1)}{w+1}{\left[  \frac{3w+1}{f_0\Big{(}2N-3(w+1)\Big{)}}\right]  }^{\frac{1}{N-1}}}.
\end{equation} 
From this expression we can see that the obtained critical point is either a center (for the cases \textbf{1} and \textbf{3}) or a saddle (for the case \textbf{2}). In the case of a center the static solution is stable. However, due to properties of a center fixed point the nearby trajectories do not converge to the static solution, but rotate around it, realising oscillating solutions. The question of how close to the static solution should a trajectory be to be trapped into infinite oscillations cannot be solved in a local analysis and requires numerical investigations, which are described in the following section.
  
    Now we find the conditions of the existence of a bounce substituting $H=0$ to (\ref{constraint1}), (\ref{Ht4}). Noticing that $T=-6H^2+\frac{6k}{a^2}=\frac{6k}{a^2}$ for $H=0$ we have
\begin{equation}
\label{abounce}
\frac{6k}{a^2}\left[ 1+f_0{\left( \frac{6k}{a^2}\right) }^{N-1}\right] =2 K\rho,
\end{equation}
\begin{equation}
\label{Htbouncem}
\dot H=\frac{k}{a^2}\left[ \frac{-3w-1+f_0{\left(\frac{6k}{a^2} \right) }^{N-1}\Big{(}(2N-3(w+1)\Big{)}}{2\left( Nf_0{\left(\frac{6k}{a^2}\right)}^{N-1}+1\right)} \right].
\end{equation}  
\\
\\Applying the conditions $\dot H>0$ and $\rho\geqslant0$ to (\ref{abounce}), (\ref{Htbouncem}) we see that the bounces exist for
\\\textbf{1.} $k=1$, ~~ $-1\leqslant f_0{\left( \frac{6k}{a^2}\right) }^{N-1}<-\frac{1}{N}$,~~ $w\in[-1; 1]$,
\\\text{~~~~}where $f_0<0$ is allowed.
\\
\\\textbf{2.} $k=1$,~~ $\frac{3w+1}{2N-3(w+1)}<f_0{\left( \frac{6k}{a^2}\right) }^{N-1}$,~~ $w\in\left[ -1; \frac{2N-3}{3}\right)$, 
\\\text{~~~~}where \textbf{a).} $f_0>0$ for $w\geqslant-\frac{1}{3}$~~ and\\
\text{\qquad\qquad~}\textbf{b).} $f_0$ can be positive as well as negative for $w<-\frac{1}{3}$.
\\
\\\textbf{3.} $k=-1$,~~ $f_0{\left( \frac{6k}{a^2}\right) }^{N-1}<\frac{3w+1}{2N-3(w+1)}$, ~~ $w\in\left(\frac{2N-3}{3}; 1\right]$,
\\\text{~~~~}where \textbf{a).} $f_0>0$ for even $N$~~ and\\
\text{\qquad\qquad~}\textbf{b).} $f_0<0$ for odd $N$.
\\ 
          
    We can see that for the positive spatial curvature bounces exist in a larger range of $w$ than the range where a static solution (stable or unstable) exists. In the case of a negative spatial curvature the conditions on the equation of state for bounce and static solutions to exist coincide for even $N$. As for the case of odd $N$, it always requires an exotic matter with $w>1$ and is not considered in the present paper.

\section{Numerical investigations}
\label{sec4}
~~~~We integrate numerically the system of three first-order differential equations: (\ref{Ht3}), (\ref{at}) and $\dot\rho=-3H(w+1)\rho$. The constraint (\ref{constraint1}) is checked at each step of the integration.

\begin{figure}[hbtp]
\includegraphics[scale=0.44]{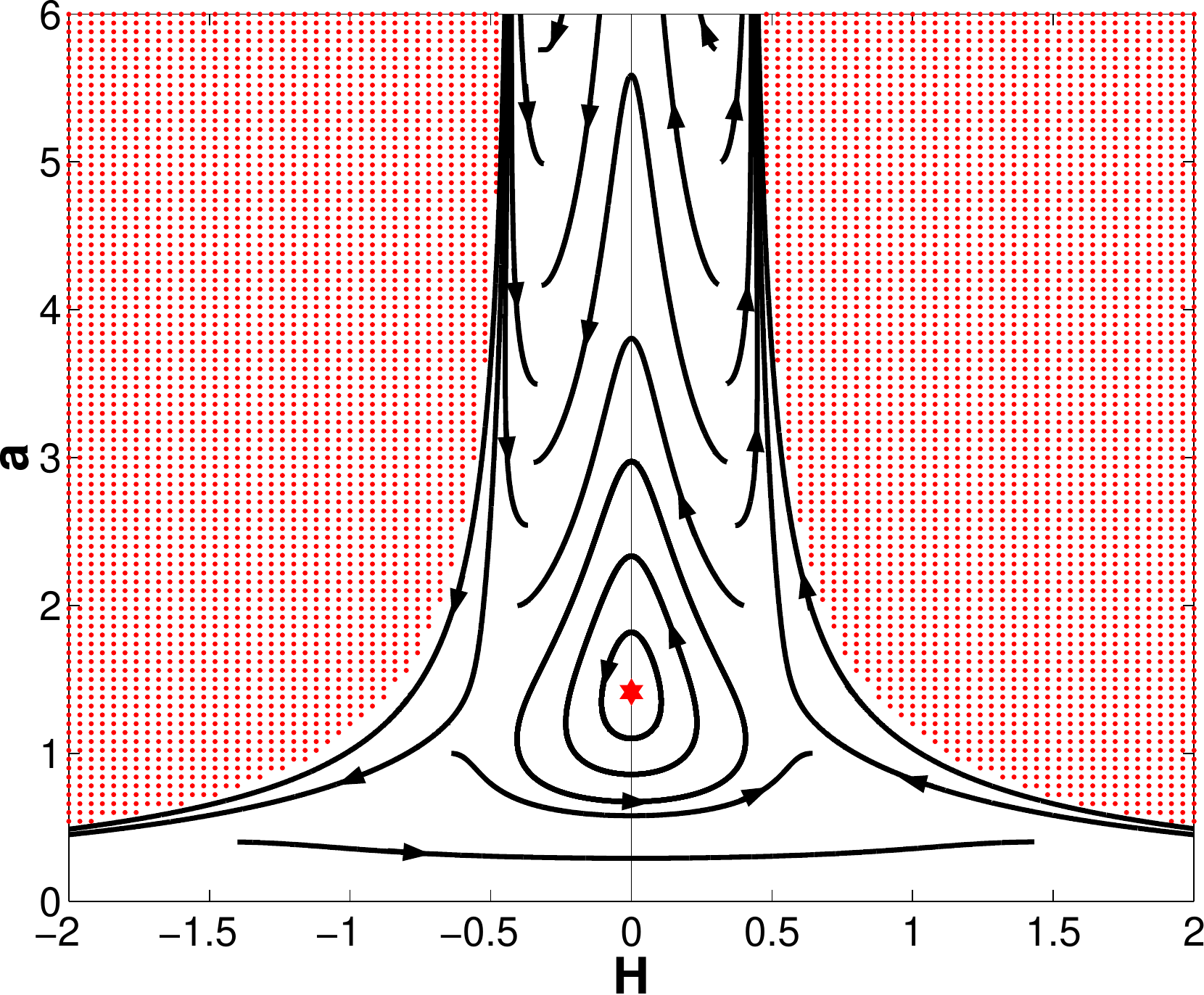}\qquad
\includegraphics[scale=0.44]{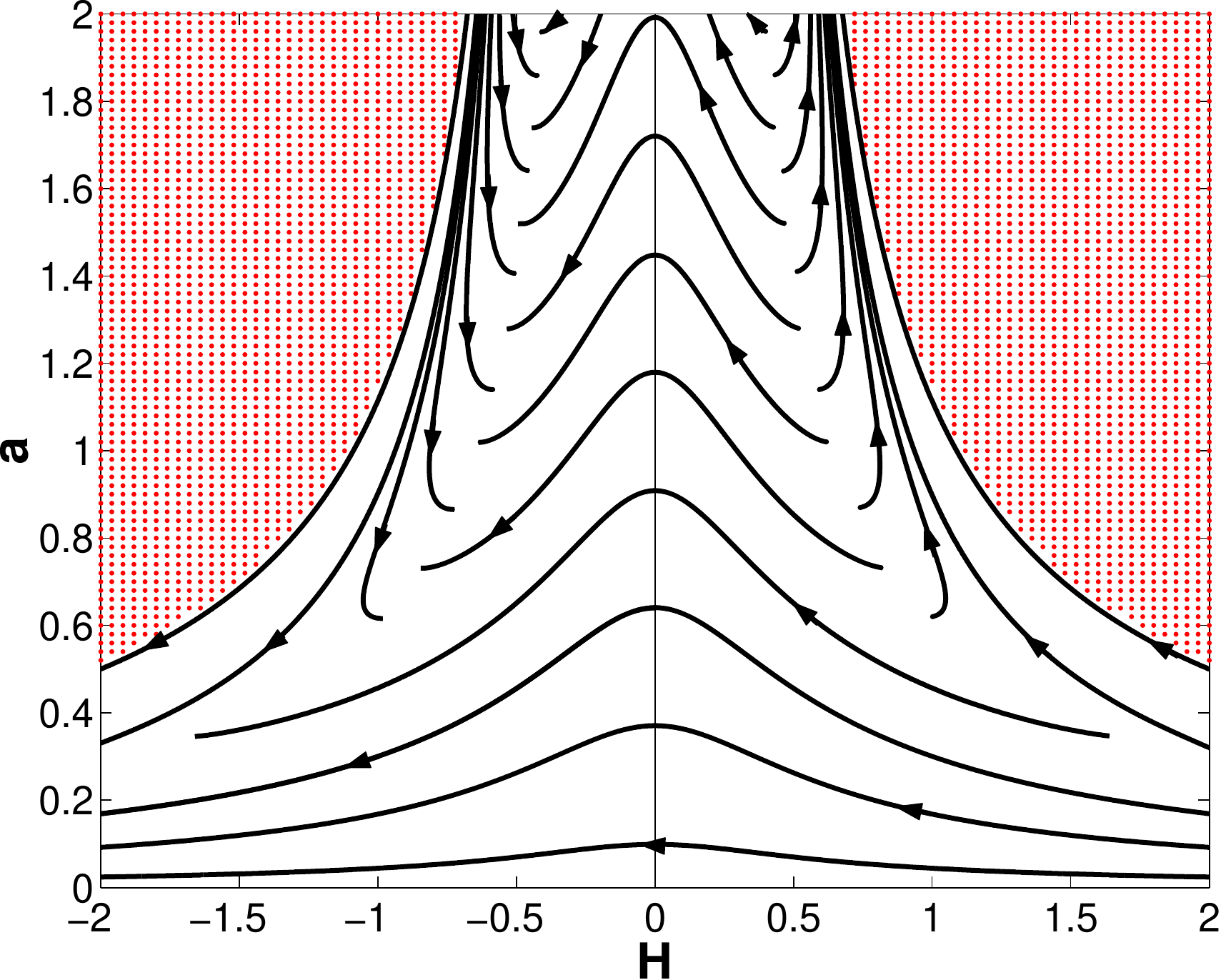}
\caption{The phase portraits are plotted for $N=2$,~~ $f_0=\frac{1}{3}$, ~~$k=1$, ~~ $w=0$ (left)~~ and ~~$w=1/3$ (right). The star denotes the stationary point. The dotted regions correspond to $\rho<0$ in the constraint equation (\ref{constraint1}).}
\label{Fig1}
\end{figure}

    The case $N=2$ and positive $f_0$ is shown in Fig.~\ref{Fig1}. The left panel corresponds to the case of $w=0$ when stable stationary solution exists. In the vicinity of this solution trajectories move around it. Note, however, that this happens only for nearby trajectories, and the fate of trajectories outside of this basin is different. If a trajectory starts from a rather big scale factor, instead of bouncing due to the presence of an $1/a^4$ term, it meets a non-standard singularity. In the non-standard singularity $\dot H$ diverges, while $H$ and $a$ are finite, so in the $a(H)$ plot it corresponds to the sudden disappearance of the phase trajectory. This means that the possibility of a bounce is not realized for a contracting Universe that is big enough initially.

    The right panel of Fig.~\ref{Fig1} presents the situation for $w=1/3$ when the stationary solution does not exist. Bouncing solutions disappeared as well. However, this does not change much the fate of a contracting large Universe, which still ends its evolution in a non-standard singularity.

\begin{figure}[hbtp]
\includegraphics[scale=0.44]{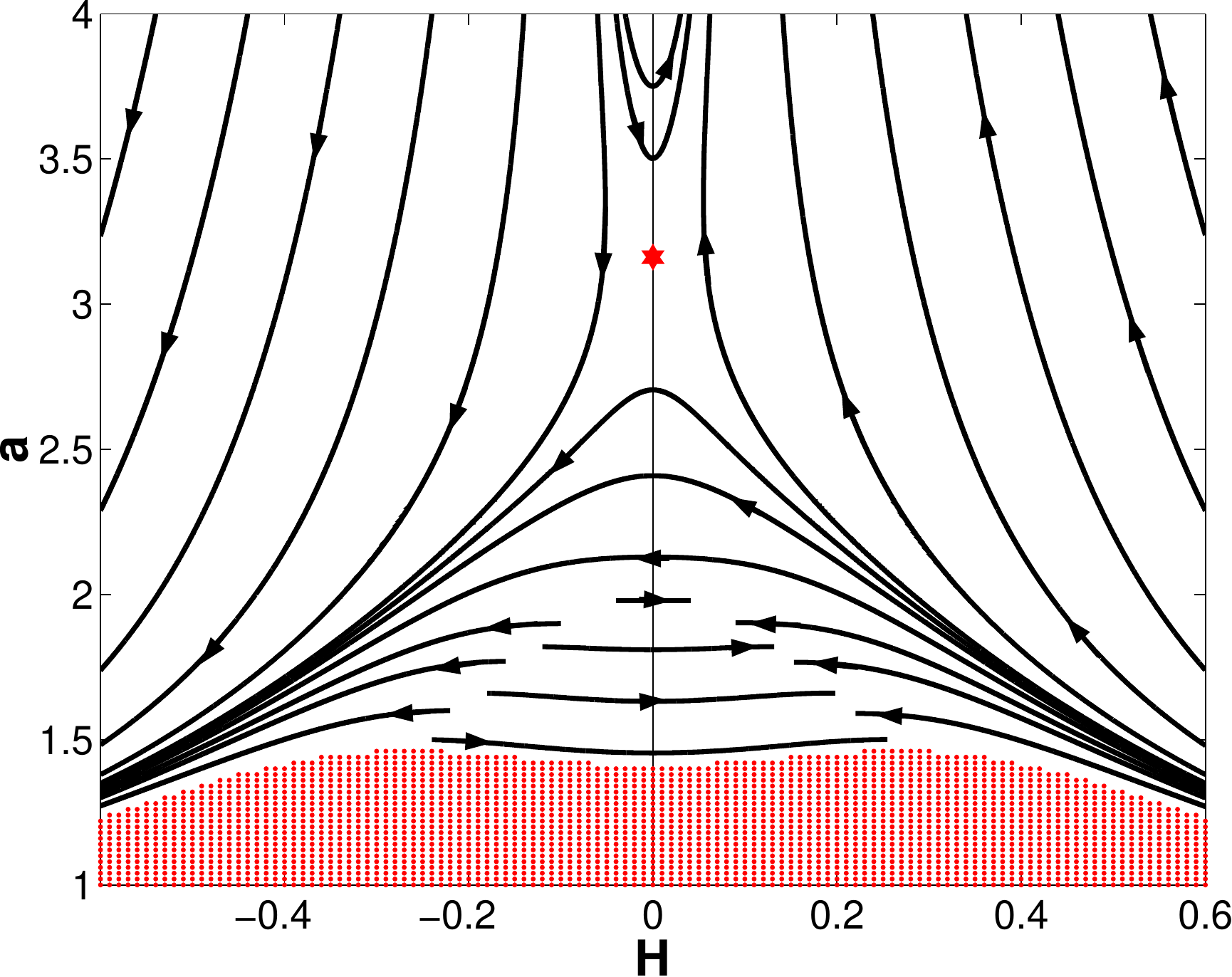}\qquad
\includegraphics[scale=0.44]{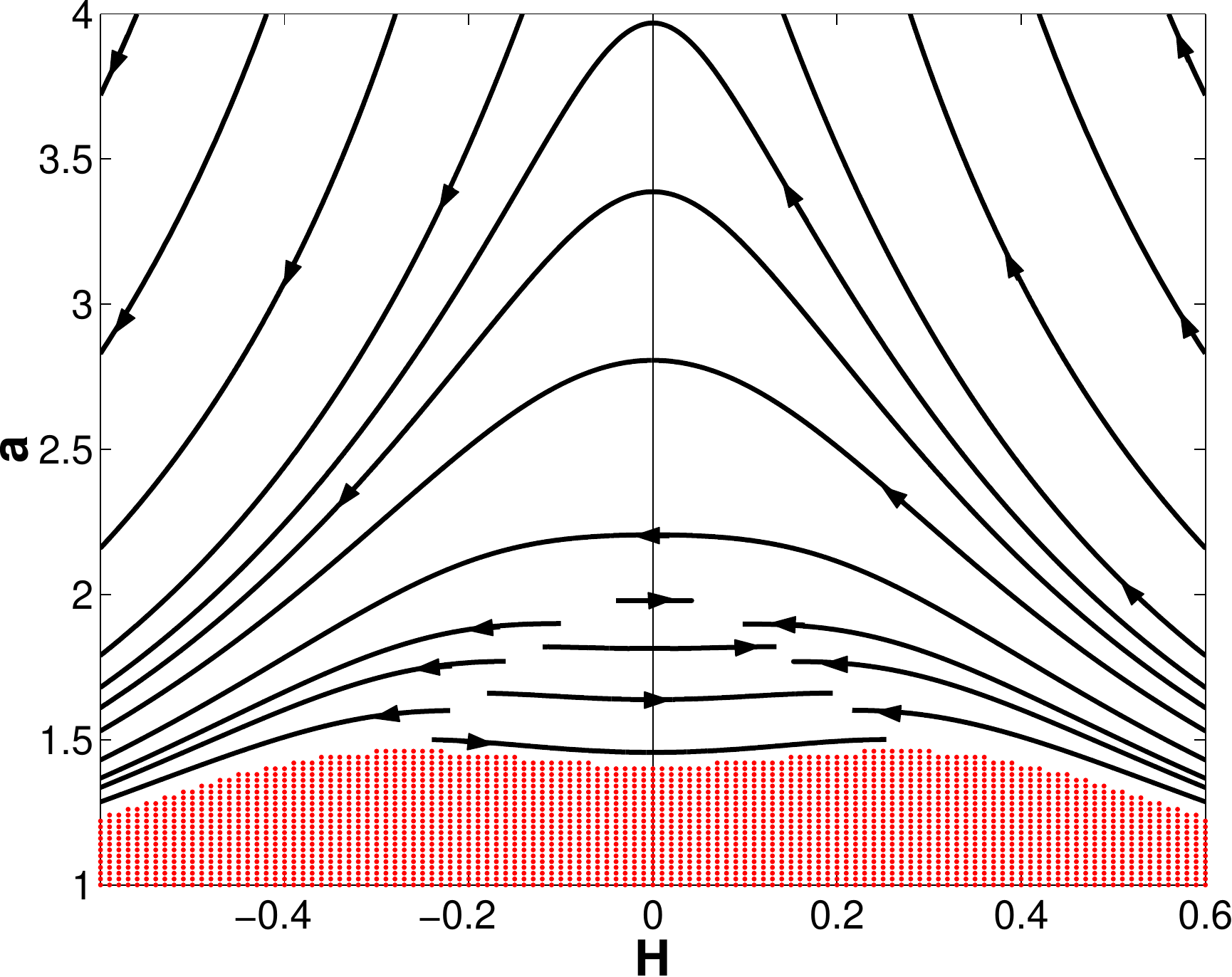}
\caption{The phase portraits are plotted for $N=2$,~~ $f_0=-\frac{1}{3}$, ~~$k=1$, ~~$w=-0.5$ (left) and $w=0$ (right). The star denotes the stationary point. The dotted regions correspond to $\rho<0$ in the constraint equation (\ref{constraint1}).}
\label{Fig2}
\end{figure}

    Negative $f_0$ leads to instability of the stationary solution if it exists (Fig.~\ref{Fig2}, left panel, $w=-0.5$). A contracting Universe either experiences a bounce or falls into a standard singularity. The right panel represents the phase portrait without the stationary solution ($w=0$). Since the conditions for a bounce are less restricted than the conditions for the stationary solution to exist, bounces are still possible. However, all trajectories with bounces begin and end in a non-standard singularity, and the fate for a big enough contracting Universe is unique --- a standard singularity.

\begin{figure}[hbtp]
\includegraphics[scale=0.44]{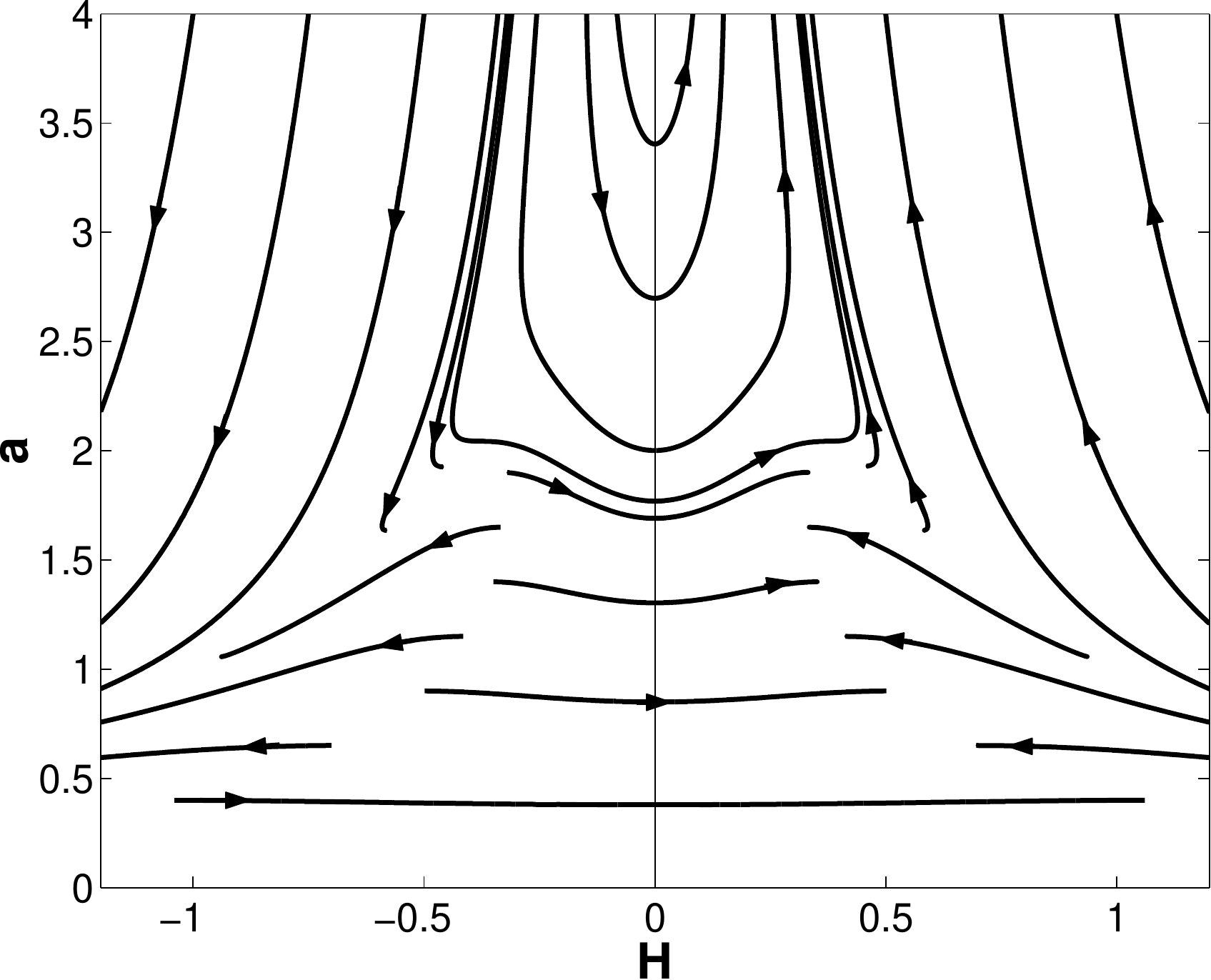}\qquad
\includegraphics[scale=0.44]{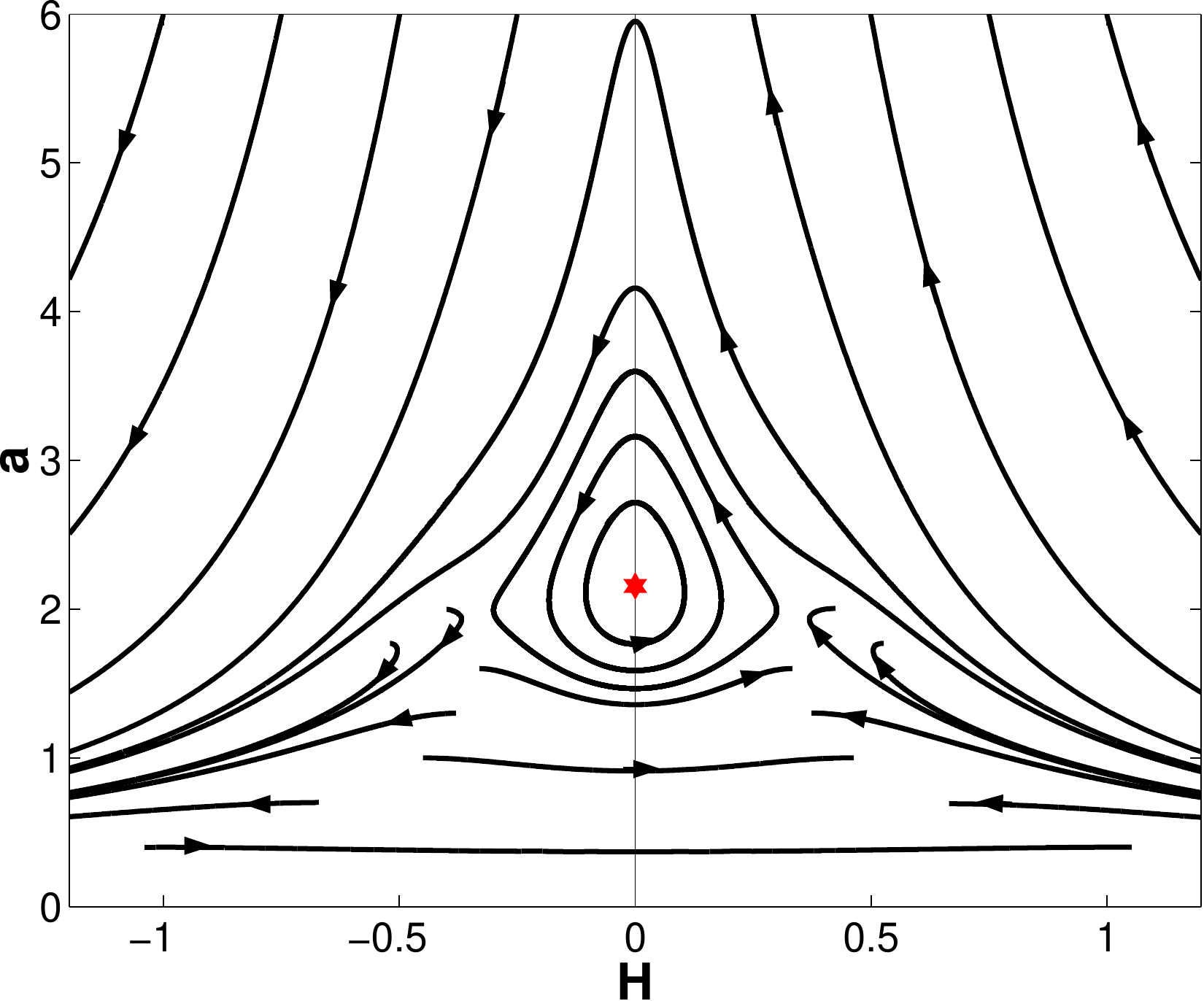}
\caption{The phase portraits are plotted for $N=3$,~~ $f_0=\frac{1}{5}$, ~~$k=1$, ~~ $w=-0.5$ (left)~~ and ~~$w=0$ (right). The star denotes the stationary point. The dotted regions correspond to $\rho<0$ in the constraint equation (\ref{constraint1}).}
\label{Fig3}
\end{figure}

    The two following plots show examples with an odd $N$. The stationary solution exists for $N=3$, $f_0>0$, $w>-1/3$ (Fig.~\ref{Fig3}, right panel). A big contracting Universe ultimately falls into a standard singularity. An example of a phase portrait without stationary solutions is shown in the left panel.

\begin{figure}[hbtp]
\includegraphics[scale=0.44]{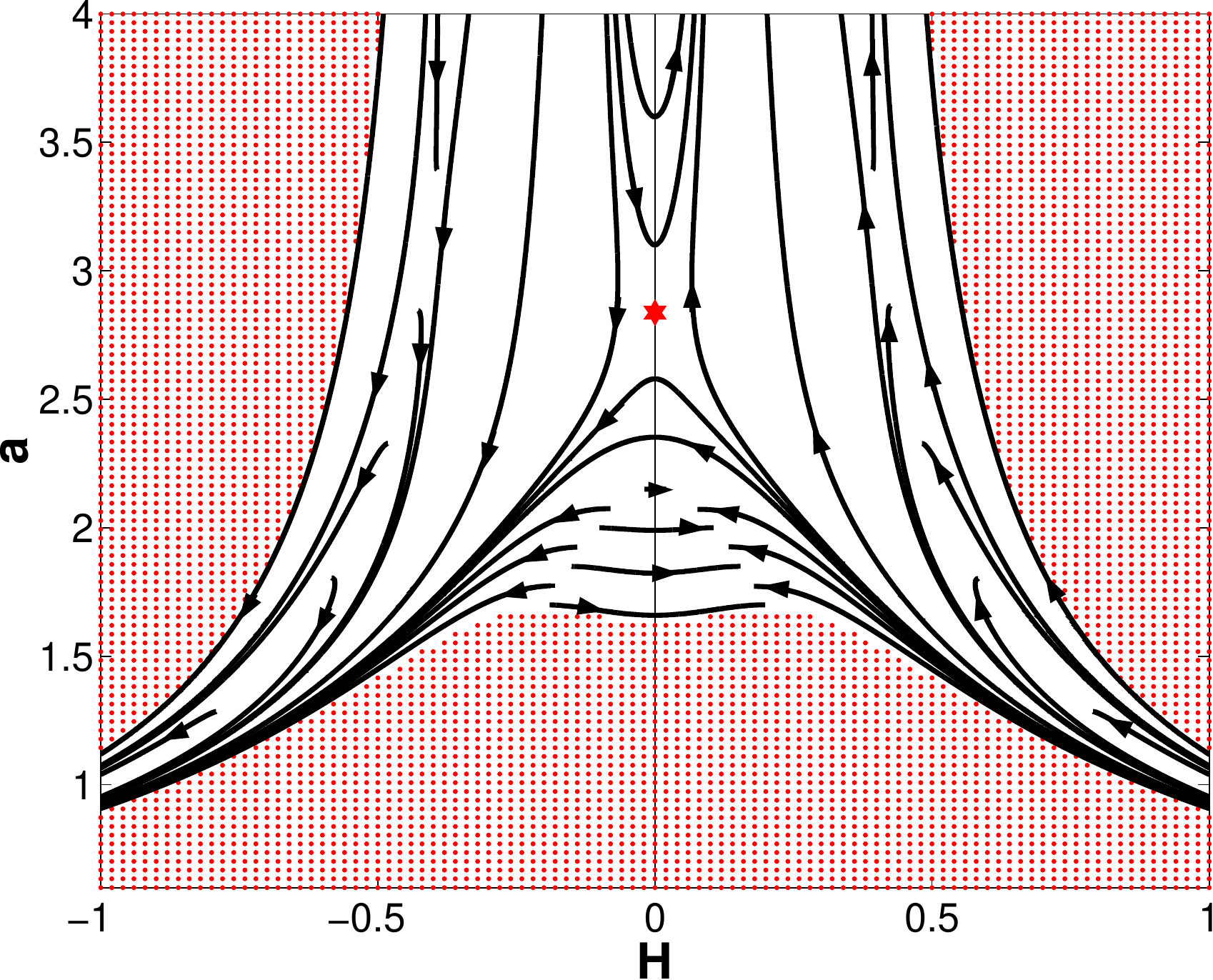}\qquad
\includegraphics[scale=0.44]{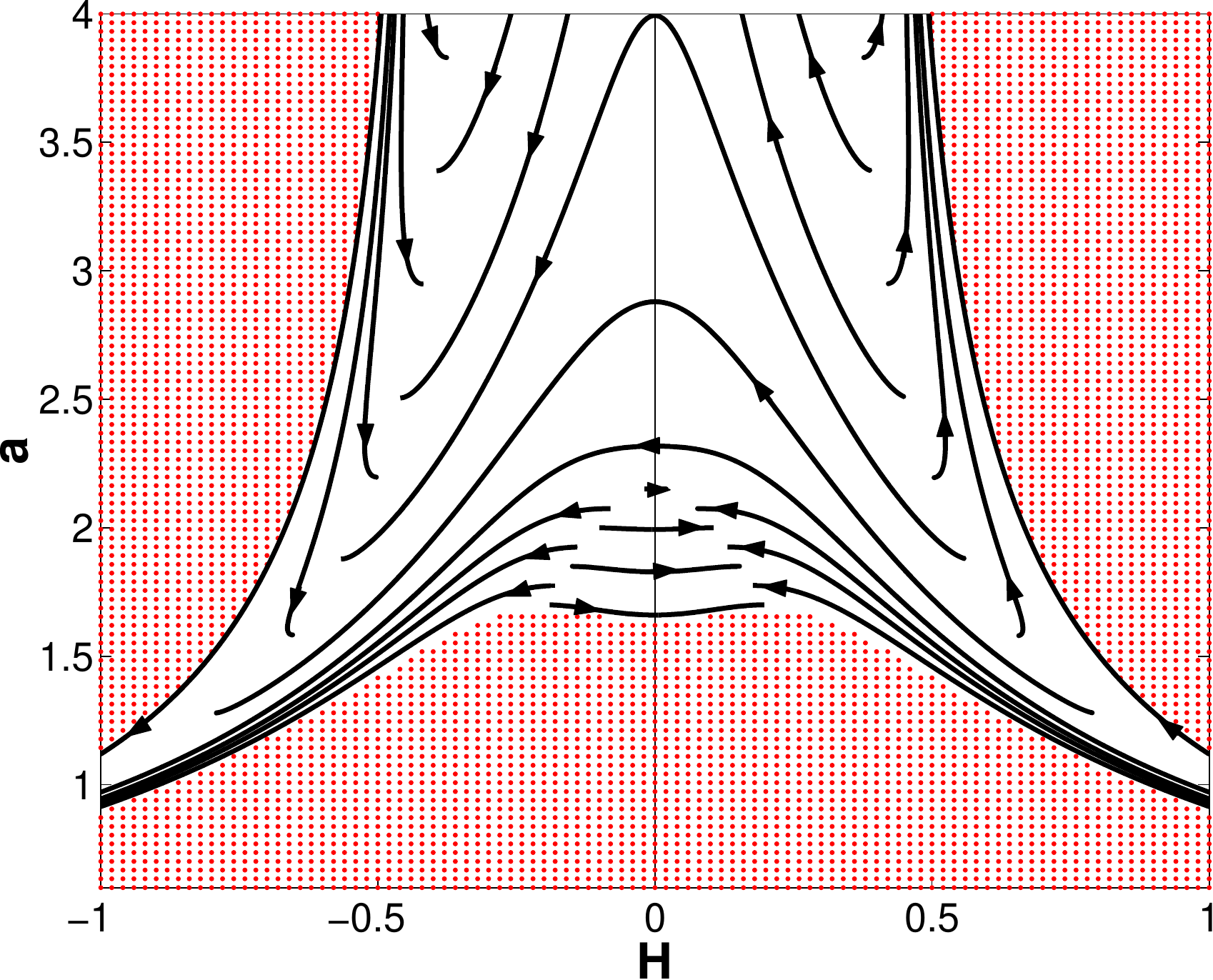}
\caption{The phase portraits are plotted for $N=3$,~~ $f_0=-\frac{1}{5}$, ~~$k=1$, ~~$w=-0.5$ (left) and $w=0$ (right). The star denotes the stationary point. The dotted regions correspond to $\rho<0$ in the constraint equation (\ref{constraint1}).}
\label{Fig4}
\end{figure}

    Negative $f_0$ leads either to an unstable stationary solution (Fig.~\ref{Fig4}, left panel) or to dynamics without stationary solutions (Fig.~\ref{Fig4}, right panel). In the former case a big enough Universe can either go through a bounce or meet a non-standard singularity, in the latter case a non-standard singularity is the unique possibility.
    
\begin{figure}[hbtp]
~~~~~~~~~~~~\includegraphics[scale=0.68]{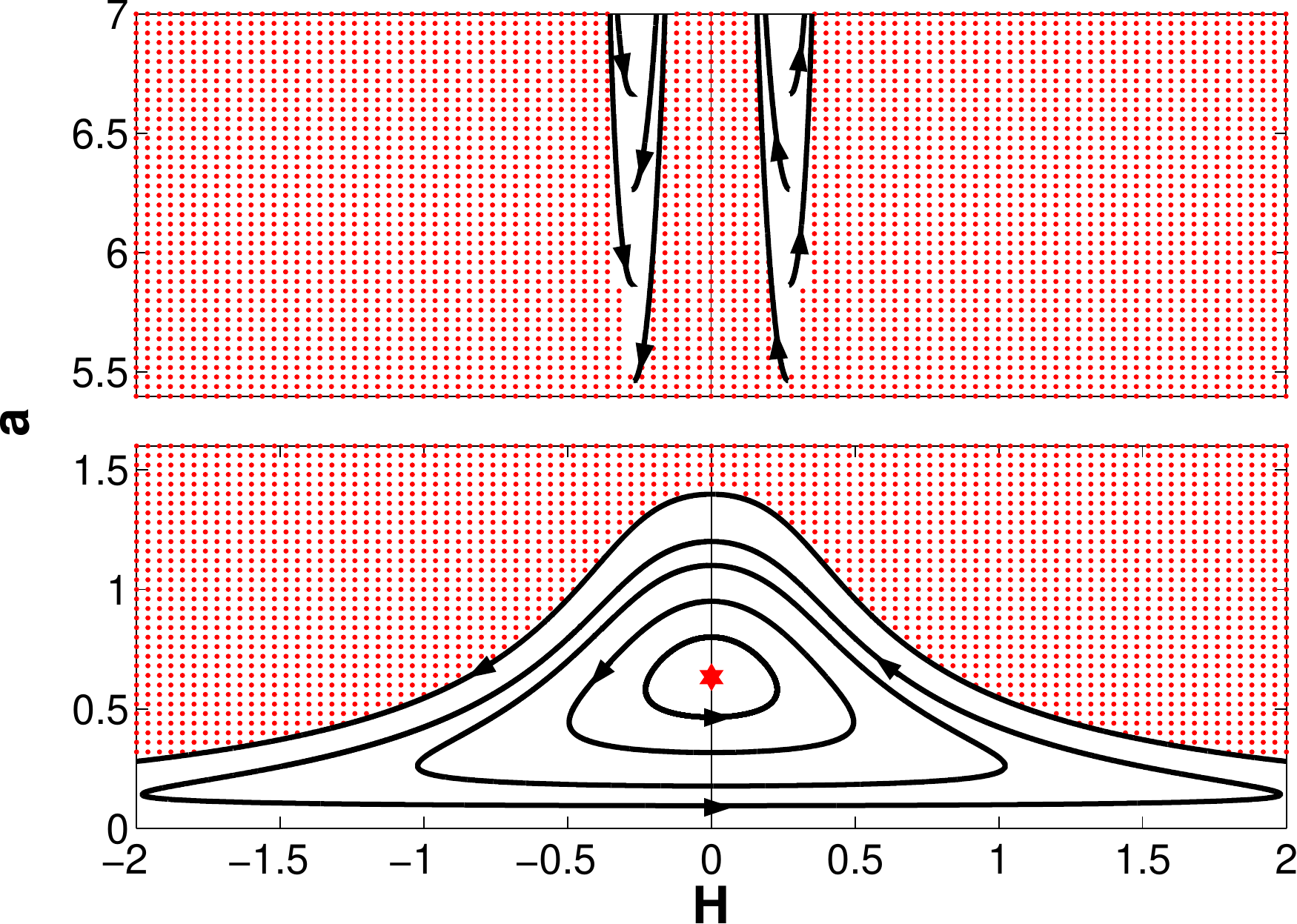}
\caption{Two parts of the phase portrait are plotted for $N=2$,~~ $f_0=\frac{1}{3}$, ~~$k=-1$, ~~ $w=0.5$. The star denotes the stationary point. The dotted zones correspond to $\rho<0$ in the constraint equation (\ref{constraint1}).}
\label{Fig5}
\end{figure}

    The phase diagram for the case of $N=2$ and a negative spatial curvature is presented in Fig.~\ref{Fig5}. It us easy to see from (\ref{constraint1}) that the maximum value of the scale factor at the point $H=0$ is equal to $a^2_{max}=6 f_0$. Larger values of the scale factor cannot correspond to its extremum. The allowed zone in the $(a, H)$ plane is divided into three separate regions. Any big contracting Universe ends its evolution in a non-standard singularity, and we have now a simple criterion for a ``big'' Universe: $a>a_{max}$.
\newpage
\section{Conclusions}
\label{sec5}
~~~~In the present paper we have considered some peculiarities of a spatially curved isotropic Universe in $f(T)$ gravity. The function $f(T)$ has been chosen in the form $f(T)=T+f_0 T^N$; the results obtained crucially depend on the parity of $N$ and the sign of $f_0$. One of the most interesting features of the dynamics is the existence of a stable static solution. For positive spatial curvature it exists for a rather wide interval of the equation of state parameter $w$, and for $N>3$ it covers the whole thermodynamically possible interval of decelerating expansion $w \in (1/3; 1]$. Moreover, for even $N$ and positive $f_0$ the stable static solution exists for a negatively curved Universe also; in this case the matter should be stiffer than an ultra-relativistic matter. 

    The conditions for possible bouncing solutions appear to be much wider than the condition $w<-1/3$ in GR. The stable static solution implies bounces, because nearby trajectories rotate around it, experiencing one bounce and one recollapse point per period. For a positive curvature bounces exist for a wider range of $w$ than the static solutions. For negative spatial curvature with an even $N$ the two conditions are the same. An odd $N$ with negative spatial curvature needs an exotic matter with $w>1$ for bounce.

    Global analysis with numerical integration of the equations of motion and constructing phase portraits show, however, that the significance of bounce solutions is less than may be thought using the results of local analysis only. All phase portraits show a common feature: any trajectory going through a bounce driven by a correction term then experiences either recollapse or a non-standard singularity rather soon. For values of the coupling constant $f_0$ of the order of unity, studied in the present paper, the maximum value of the scale factor is of the same order of magnitude as the scale factor in the bounce point. On the other side, this means that a contracting Universe with large enough initial scale factor can go through a bounce only if $w$ lies within the GR allowed interval $w<-1/3$. For stiffer matter bouncing trajectories cannot reach the low-curvature regime of a relatively big Universe. Further studies are needed to correctly quantify this qualitative result and determine how the maximum value of the scale factor after a bounce depends on $f_0$. Currently, it seems highly unlikely that new bounce solutions can be important for the future of our ``big'' Universe if it starts at some time to contract.

    For the negative curvature case the structure of the phase space is different and allows for a simple analytical expression of the maximal scale factor after bounce. The phase source is divided into three disconnected zones, one zone in the small scale factor range where trajectories move around static solution, and two symmetric zones in the large scale factor range where evolution of scale factor is monotonous. Since all bounces are located in the first zone, a Universe large enough to be located in the second zone cannot experience any bounce. The maximum scale factor of the first zone is $a_{max}=\sqrt{6 f_0}$, giving an upper bound for the scale factors of bouncing trajectories.

\section*{Acknowledgements}
~~~~Maria A. Skugoreva and Alexey V. Toporensky are supported by RSF Grant \textnumero 16-12-10401 and Alexey V. Toporensky is supported by the Russian Government Program of Competitive Growth of Kazan Federal University.

\end{document}